\documentclass[11pt,superscriptaddress,aps,prd,preprint]{revtex4}
\usepackage{amsmath}

\makeatletter
\usepackage[T1]{fontenc}
\usepackage{amsmath}
\usepackage{amssymb}
\usepackage{graphicx}

\usepackage{slashed}

\newcommand{\bea}{\begin{eqnarray}}
\newcommand{\eea}{\end{eqnarray}}

\begin{document}

\title{Casimir effect and Stefan-Boltzmann law in Yang-Mills theory at finite temperature}

\author{A. F. Santos}\email{alesandroferreira@fisica.ufmt.br}
\affiliation{Instituto de F\'{\i}sica, Universidade Federal de Mato Grosso,\\
78060-900, Cuiab\'{a}, Mato Grosso, Brazil}

\author{Faqir C. Khanna\footnote{Professor Emeritus - Physics Department, Theoretical Physics Institute, University of Alberta\\
Edmonton, Alberta, Canada}}\email[]{khannaf@uvic.ca}
\affiliation{Department of Physics and Astronomy, University of Victoria,\\-
3800 Finnerty Road Victoria, BC, Canada}

\begin{abstract}

A non-Abelian gauge theory describes the strong interactions among particles with the commutator of generators are non-zero. An $SU(3)$ gauge theory describes the interactions that lead to nuclear forces among particles. The Lagrangian density refers to fermions with colour and flavour and the gauge field quanta implying gluons. The gauge theory is treated at finite temperature using the Thermo Field Dynamics (TFD). Using self interaction of gluons the Stefan-Boltzmann law and the Casimir effect are calculated at finite temperature. An appendix is attached to give a response of a massless quarks in gauge theory.

\end{abstract}

\maketitle

\section{Introduction}

The Gauge theories describe three of the four fundamental forces of nature: the electromag- netism, the weak force and the strong force. These are described under the name, the standard model. The Quantum Electrodynamics (QED), that explains the quantum effects of electromagnetism, is a gauge theory which is verified to a high precision and is explained as a $U(1)$ Abelian gauge theory. Yang and Mills \cite{YM} then generalized this Abelian gauge theory to the Non-Abelian gauge theory that proved successful in the formulation of both Electroweak Unification and Quantum Chromodynamics (QCD). The electroweak interaction is described by an $SU(2)\times U(1)$ group and  while $SU(3)$ group satisfies QCD. One of the major differences between Abelian and non-Abelian gauge theories is the existence of self interactions in the non-Abelian case. QCD is a quantum field theory that describes the strong interaction, between quarks and gluons, with non-Abelian gauge fields \cite{Marciano, Marciano1, Brambilla}. QCD is an asymptotically free theory which implies that the interaction between quarks and gluons becomes weak at high energies and exhibits confinement. The Lagrangian of QCD is composed of quarks and gluons fields. Here the study is developed about presence of the Casimir effect and the Stefan- Boltzmann law for the pure Yang-Mills theory. This paper deals with the gluons field only. In this theory the gluons interact among themselves, then a natural question is: how these self-interactions lead to the Stefan-Boltzmann law and the Casimir effect at zero and finite temperature?

The Casimir effect was predicted by H. Casimir \cite{Casimir}. This effect was measured when two parallel conducting plates are attracted due to vacuum fluctuations of the electromagnetic field or other topological effects \cite{Milton, Milonni, Plunien, Bordag}. The first experimental confirmation of the Casimir effect was carried out by Sparnaay \cite{Sparnaay}. Subsequent experiments have established this effect to a high degree of accuracy \cite{Lamoreaux, Mohideen}. This phenomenon was first predicted and experimentally confirmed for the electromagnetic field, however it is likely to appear for any other quantum field. The Casimir effect for the gravitational field has been studied \cite{Quach, GEM}. Role of Lorentz violation for Yang-Mills theory has been investigated \cite{LV1, LV2, LV3}. For non-Abelian gauge theory using first-principle numerical simulations have been considered \cite{Chernodub}, among others. Corrections due to the finite temperature have been calculated for electromagnetic field \cite{Khanna0}, Kalb-Ramond field \cite{Ademir}, gravitational field \cite{GEM} including the role of Lorentz-violation in these effects cite{LV1, LV2, LV3}. The Casimir effect and Stefan-Boltzmann law for the Yang-Mills field are investigated at finite temperature for a gluon field. The finite temperature effects are calculated using the Thermo Field Dynamics (TFD). In the Appendix A, thermal effects are considered for a quarks and are calculated using Thermo Field Dynamics.

Temperature effects in a quantum field theory may be introduced using three different, but equivalent, formalisms. These formalisms are: (i) the Matsubara formalism or the imaginary time formalism \cite{Matsubara}; (ii) the closed time path formalism \cite{Schwinger} and (iii) the Thermo Field Dynamics (TFD) formalism \cite{Umezawa1, Umezawa2, Umezawa22, Khanna1, Khanna2}. Here the TFD formalism is used to introduce temperature effects in the Casimir effect and Stefan-Boltzmann Law for the gauge theory. TFD, a real-time finite temperature formalism, is a thermal quantum field theory that depends on the doubling of the original Fock space, composed of the original and a fictitious space (tilde space), using the Bogoliubov transformation. The original and tilde space are related by a mapping, tilde conjugation rules. The Bogoliubov transformation is a rotation involving the two spaces, original and tilde, which introduce the temperature effects.

This paper is organized as follows. In section II, the Lagrangian that describes the Yang-Mills field is presented and the corresponding energy-momentum tensor is determined. In section III, a brief introduction to the TFD formalism is considered. In section IV, the Stefan-Boltzmann law at zero and finite temperature for the Gauge theory is calculated. In section V, the Casimir effect at zero and finite temperature for the non-Abelian theory are investigated. In section VI, some concluding remarks are presented. In Appendix A, the Stefan-Boltzmann law and the Casimir effect at zero and finite temperature are calculated for the non-interacting massless QCD.

\section{The energy-momentum tensor of gauge theory}

The gauge theory with a non-Abelian symmetry group is used to calculate the Casimir effect and Stefan-Boltzmann Law. Its interaction Lagrangian density for the Gluon field is given as
\bea
{\cal L}=-\frac{1}{4}F_{\mu\nu}^a F^{\mu\nu a},
\eea
with
\bea
F_{\mu\nu}^a=\partial_\mu A_\nu^a-\partial_\nu A_\mu^a+gf^{abc}A_\mu^b A_\nu^c \label{FF}
\eea
which is the field strength tensor of the gluon (non-Abelian) field $A_\mu^a$, $g$ is the coupling constant and $f^{abc}$ are structure constants of the Lie algebra. The  index $a$ is summed over the generators of the gauge group and for a SU(N) group one has $ a,b,c=1\cdots N^{2}-1.$ 

The energy-momentum tensor associated to the non-Abelian field is 
\bea
T^{\mu\nu}=-F^{\mu\alpha a}F^{a\nu}\,_\alpha+\frac{1}{4}\eta^{\mu\nu}F^a_{\alpha\rho}F^{\alpha\rho a},
\eea
with $\eta^{\mu\nu}=(+,-,-,-)$ being the Minkowski space-time. 

Our main objective in this paper is to calculated the Stefan-Boltzmann law and the Casimir effect at finite temperature for the gluon field in the Lagrangian density for Yang-Mills. Then the vacuum expectation value of the energy-momentum tensor needs to be determined. Here the energy-momentum tensor is written so as to avoid a product of field operators at the same space-time point. Then the energy-momentum tensor is
\bea
T^{\mu\nu}(\omega)=\lim_{x'\rightarrow \omega}\tau\left[-F^{\mu\alpha a}(\omega)F^{a\nu}\,_\alpha(x')+\frac{1}{4}\eta^{\mu\nu}F^a_{\alpha\rho}(\omega)F^{\alpha\rho a}(x')\right],\label{EMT}
\eea
where $\tau$ is the time ordering operator and $x'=x, y, z$. For simplicity define
\bea
{\cal F}^{\mu\nu a, \alpha\rho a}(\omega,x')&\equiv&\tau\left[F^{\mu\nu a}(\omega)F^{\alpha\rho a}(x')\right]\nonumber\\
&=&F^{\mu\nu a}(\omega)F^{\alpha\rho a}(x')\theta(\omega_0-x_0')+F^{\alpha\rho a}(x')F^{\mu\nu a}(\omega)\theta(\omega_0'-x_0).
\eea
After some algebra, this leads to
\bea
{\cal F}^{\mu\nu a, \alpha\rho a}(\omega,x')={\cal F}_1^{\mu\nu a, \alpha\rho a}(\omega,x')+{\cal F}_2^{\mu\nu a, \alpha\rho a}(\omega,x')+{\cal F}_3^{\mu\nu a, \alpha\rho a}(\omega,x'),
\eea
where
\bea
{\cal F}_1^{\mu\nu a, \alpha\rho a}(\omega,x')&=&\Gamma^{\mu\nu,\alpha\rho,\lambda\epsilon}\,\tau\left[A^a_\lambda(\omega)A^a_{\epsilon}(x)\right]+gf^{abc}\,\Gamma^{\mu\nu}_\lambda\,\tau\left[A^{a\lambda}(\omega)A^{b\alpha}(x)A^{c\rho}(y)\right],\\
{\cal F}_2^{\mu\nu a, \alpha\rho a}(\omega,x')&=&I^{\nu a,\alpha\rho a}(\omega,x')\,n_0^\mu\,\delta(\omega_0-x_0')-I^{\mu a,\alpha\rho a}(\omega,x')\,n_0^\nu\,\delta(\omega_0-x_0'),\\
{\cal F}_3^{\mu\nu a, \alpha\rho a}(\omega,x')&=&gf^{abc}\,\Gamma'^{\alpha\rho}_\lambda\tau\left[A^{b\mu}(\omega)A^{c\nu}(x)A^{a\lambda}(y)\right]\nonumber\\
&+&g^2f^{abc} f^{ade}\tau\left[A^{b\mu}(\omega)A^{c\nu}(x)A^{d\alpha}(y)A^{e\rho}(z)\right],
\eea
with
\bea
\Gamma^{\mu\nu,\alpha\rho,\lambda\epsilon}&=&\left(g^{\nu\lambda}\partial^\mu-g^{\mu\lambda}\partial^\nu\right)\left(g^{\rho\epsilon}\partial'^\alpha-g^{\alpha\epsilon}\partial'^\rho\right)\\
\Gamma^{\mu\nu}_\lambda&=&g^\nu\,_\lambda\partial^\mu-g^\mu\,_\lambda\partial^\nu\\
I^{\nu a,\alpha\rho a}(\omega,x')&=&\left[A^{a\nu}(\omega), F^{a\alpha\rho}(x')\right]\label{eq12}\\
\partial^\mu\theta(\omega_0-x_0')&=&n^\mu_0\delta(\omega_0-x_0')
\eea
and $n^\mu_0=(1,0,0,0)$ is a time-like vector.

The quantization of the non-Abelian gauge fields requires the construction of the canonical momenta
\bea
\pi^{\mu a}=\frac{\partial{\cal L}}{\partial(\partial_0 A^a_\mu)}=-F^{0\mu,a},
\eea
with components
\bea
\pi^{0 a}=0\quad\quad\quad{\rm and}\quad\quad\quad\pi^{i a}=-F^{0i, a}=E^{i a}
\eea
here $E^{i a}$ is the non-Abelian field. Then the standard canonical commutation relation is
\bea
\left[A^a_i(\omega),\pi^b_j(x')\right]=i\delta^{ab}\,\delta_{ij}\,\delta^3(\vec{\omega}-\vec{x'}).
\eea
All other commutation relations are zero. Thus eq. (\ref{eq12}) becomes
\bea
I^{\nu a,\alpha\rho a}(\omega,x')=in^\alpha_0\,\delta^{aa}\,\delta^{\nu\rho}\,\delta^3(\vec{\omega}-\vec{x'})-in^\rho_0\,\delta^{aa}\,\delta^{\nu\rho}\,\delta^3(\vec{\omega}-\vec{x'}).
\eea
Then 
\bea
{\cal F}^a\,_{\alpha\rho,}\,^{\alpha\rho a}(\omega,x,y,z)&=&\Gamma_{\alpha\rho,}\,^{\alpha\rho,\lambda\epsilon}\,\tau\left[A^a_\lambda(\omega)A^a_{\epsilon}(x)\right]+g f^{abc}\Bigl(\Gamma_{\alpha\rho\lambda}\,\tau\left[A^{a\lambda}(\omega)A^{b\alpha}(x)A^{c\rho}(y)\right]\nonumber\\
&+&\Gamma'^{\alpha\rho}\,_\lambda\,\tau\left[A^b_\alpha(\omega)A^c\rho(x)A^{a\lambda}(y)\right]\Bigl)+g^2f^{abc}f^{ade}\tau\left[A^b_\alpha(\omega)A^c_\rho(x)A^{d\alpha}(y)A^{e\rho}(z)\right]\nonumber\\
&+&I^a_{\rho,}\,^{\alpha\rho a}\,n_{\alpha 0}\delta(\omega_0-x'_0)-I^a_{\alpha,}\,^{\alpha\rho a}\,n_{\rho 0}\delta(\omega_0-x'_0),
\eea
and the energy-momentum tensor for the gluon field, eq.(\ref{EMT}), becomes
\bea
T^{\mu\nu}(\omega)&=&\lim_{x,y,z\rightarrow \omega}\Bigl\{-\Delta^{\mu\nu,\lambda\epsilon}\,\tau\left[A^a_\lambda(\omega)A^a_{\epsilon}(x)\right]+2gf^{abc}\Delta^{\mu\nu}\,_{\lambda\delta\Lambda}\,\tau\left[A^{a\lambda}(\omega)A^{b\delta}(x)A^{c\Lambda}(y)\right]\nonumber\\
&+&g^2f^{abc}f^{ade}\Delta^{\mu\nu}\,_{\lambda\Lambda\delta\rho}\,\tau\left[A^{b\lambda}(\omega)A^{c\Lambda}(x)A^{d\delta}(y)A^{e\rho}(z)\right]\nonumber\\
&-&i\left(n^\mu_0 n^\nu_0-n_{\alpha 0}n^\alpha_0\delta^{\mu\nu}\right)\delta^4(\omega-x')\Bigl\},
\eea
where
\bea
\Delta^{\mu\nu,\lambda\epsilon}&=&\Gamma^{\mu\alpha,\nu}\,_\alpha\,^{,\lambda\epsilon}-\frac{1}{4}\eta^{\mu\nu}\,\Gamma_{\alpha\rho,}\,^{\alpha\rho,\lambda\epsilon}\\
\Delta^{\mu\nu}\,_{\lambda\delta\Lambda}&=&-\Gamma^{\mu}\,_{\Lambda\lambda}\,g^\nu\,_\delta+\frac{1}{4}\eta^{\mu\nu}\,\Gamma_{\delta\Lambda\lambda}\\
\Delta^{\mu\nu}\,_{\lambda\delta\Lambda\rho}&=&-g^\mu\,_\lambda g^\nu\,_\delta g_{\Lambda\rho}+\frac{1}{4}\eta^{\mu\nu}g_{\Lambda\rho}g_{\lambda\delta}.
\eea

The vacuum expectation value of the energy-momentum tensor for the gluon field is
\bea
\bigl\langle 0| T^{\mu\nu}(\omega)|0\bigl\rangle&=&\bigl\langle T^{\mu\nu}(\omega)\bigl\rangle\nonumber\\
&=&\lim_{x,y,z\rightarrow \omega}\Bigl\{-\Delta^{\mu\nu,\lambda\epsilon}\,\Bigl\langle 0\Bigl|\tau\left[A^a_\lambda(\omega)A^a_{\epsilon}(x)\right]\Bigl|0\Bigl\rangle\nonumber\\
&+&2gf^{abc}\Delta^{\mu\nu}\,_{\lambda\delta\Lambda}\,\Bigl\langle 0\Bigl|\tau\left[A^{a\lambda}(\omega)A^{b\delta}(x)A^{c\Lambda}(y)\right]\Bigl|0\Bigl\rangle\nonumber\\
&+&g^2f^{abc}f^{ade}\Delta^{\mu\nu}\,_{\lambda\Lambda\delta\rho}\,\Bigl\langle 0\Bigl|\tau\left[A^{b\lambda}(\omega)A^{c\Lambda}(x)A^{d\delta}(y)A^{e\rho}(z)\right]\Bigl|0\Bigl\rangle\nonumber\\
&-&i\left(n^\mu_0 n^\nu_0-n_{\alpha 0}n^\alpha_0\delta^{\mu\nu}\right)\delta^4(\omega-x')\langle 0|0\rangle\Bigl\},\label{VEV}
\eea
where
\bea
\Bigl\langle 0\Bigl|\tau\left[A^a_\lambda(\omega)A^a_{\epsilon}(x)\right]\Bigl|0\Bigl\rangle&=&\delta^{ab}\Bigl\langle 0\Bigl|\tau\left[A^a_\lambda(\omega)A^b_{\epsilon}(x)\right]\Bigl|0\Bigl\rangle\nonumber\\
&=&i\delta^{ab}D^{ab}_{\lambda\epsilon}(\omega-x)
\eea
with
\bea
D^{ab}_{\lambda\epsilon}(\omega-x)=\delta^{ab}\,\eta_{\lambda\epsilon}\,G_0(\omega-x)
\eea
being the gluon propagator, that is described in FIG.1, and $G_0(\omega-x)$ is the massless scalar field propagator given by
\bea
G_0(\omega-x)=-\frac{i}{(2\pi)^2}\frac{1}{(\omega-x)^2-i\epsilon}.\label{G0}
\eea
The second and third terms in eq. (\ref{VEV}) are associated with the self-interactions of gluons. 
\begin{figure}[h]
\includegraphics[scale=0.6]{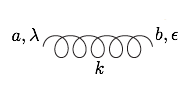}
\caption{Gluon propagator in momentum space.}
\end{figure}
The three-gluon vertex in coordinate space \cite{Leupold} is defined as
\bea
\Bigl\langle 0\Bigl|\tau\left[A^{a\lambda}(\omega)A^{b\delta}(x)A^{c\Lambda}(y)\right]\Bigl|0\Bigl\rangle=-ig f^{abc}G_3^{\lambda\delta\Lambda}(\omega, x, y),
\eea
where
\bea
G_3^{\lambda\delta\Lambda}(\omega, x, y)&=&\eta^{\lambda\delta}\left(\partial_\omega^\Lambda-\partial_x^\Lambda\right)\delta(x-y)\delta(\omega-y)\nonumber\\
&+&\eta^{\delta\Lambda}\left(\partial_x^\lambda-\partial_y^\lambda\right)\delta(y-\omega)\delta(x-\omega)\nonumber\\
&+&\eta^{\Lambda\lambda}\left(\partial_y^\delta-\partial_\omega^\delta\right)\delta(\omega-x)\delta(y-x),
\eea
and the four-gluon vertex is
\bea
\Bigl\langle 0\Bigl|\tau\left[A^{b\lambda}(\omega)A^{c\Lambda}(x)A^{d\delta}(y)A^{e\rho}(z)\right]\Bigl|0\Bigl\rangle=-ig^2\, G_4^{bcde, \lambda\Lambda\delta\rho}(\omega, x, y, z)
\eea
with
\bea
G_4^{bcde, \lambda\Lambda\delta\rho}(\omega, x, y, z)&=&\Bigl[f^{bcf}f^{def}\left(\eta^{\lambda\delta}\eta^{\rho\Lambda}-\eta^{\lambda\rho}\eta^{\Lambda\delta}\right)+f^{bdf}f^{cef}\left(\eta^{\lambda\rho}\eta^{\Lambda\delta}-\eta^{\lambda\Lambda}\eta^{\delta\rho}\right)\nonumber\\
&+&f^{bef}f^{cdf}\left(\eta^{\lambda\Lambda}\eta^{\delta\rho}-\eta^{\lambda\delta}\eta^{\rho\Lambda}\right)\Bigl]\delta(z-\omega)\delta(z-x)\delta(z-y).
\eea
The Feynman diagrams that describe the 3-gluon and 4-gluon vertices are shown in FIG.2. 
\begin{figure}[h]
\includegraphics[scale=0.5]{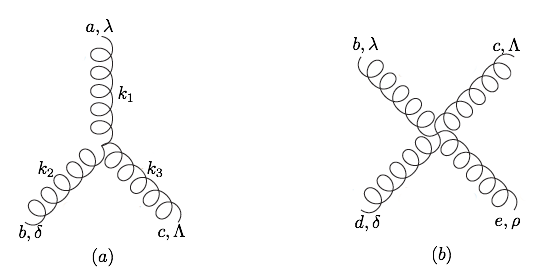}
\caption{Feynman rules in momentum space for the tree-level vertices with 3-gluon (a) and 4-gluon (b). All momenta are flowing into the vertex.}
\end{figure}
Then
\bea
\bigl\langle T^{\mu\nu}(\omega)\bigl\rangle&=&-i\lim_{x,y,z\rightarrow \omega}\Bigl\{8\Delta^{\mu\nu,\lambda\epsilon}\,g_{\lambda\epsilon}G_0(\omega-x)+2g^2f^{abc}f^{abc}\Delta^{\mu\nu}\,_{\lambda\delta\Lambda}\,G_3^{\lambda\delta\Lambda}(\omega, x, y)\nonumber\\
&+&g^4f^{abc}f^{ade}\Delta^{\mu\nu}\,_{\lambda\Lambda\delta\rho}\,G_4^{bcde, \lambda\Lambda\delta\rho}(\omega, x, y, z)\nonumber\\
&-&i\left(n^\mu_0 n^\nu_0-n_{\alpha 0}n^\alpha_0\delta^{\mu\nu}\right)\delta^4(\omega-x')\langle 0|0\rangle\Bigl\}.\label{VEV2}
\eea
Now different components are calculated separately as
\bea
\Delta^{\mu\nu,\lambda\epsilon}\,g_{\lambda\epsilon}&=&2\left(\partial^\mu_\omega\partial^\nu_x-\frac{1}{4}\eta^{\mu\nu}\partial^\alpha_\omega\partial_{x\alpha}\right)\\
\Delta^{\mu\nu}\,_{\lambda\delta\Lambda}\,G_3^{\lambda\delta\Lambda}(\omega, x, y)&=&\Bigl[-\partial^\mu_\omega\left(\partial^\nu_\omega-\partial^\nu_x\right)+\eta^{\mu\nu}\partial_{\omega\Lambda}\left(\partial^\Lambda_\omega-\partial^\Lambda_x\right)\nonumber\\
&-&\frac{3}{4}\eta^{\mu\nu}\partial_{x\Lambda}\left(\partial^\Lambda_\omega-\partial^\Lambda_x\right)\Bigl]\delta(x-y)\delta(\omega-y)\nonumber\\
&+&\Bigl[-\partial^\mu_\omega\left(\partial^\nu_x-\partial^\nu_y\right)+\partial^\nu_\omega\left(\partial^\mu_x-\partial^\mu_y\right)\Bigl]\delta(y-\omega)\delta(x-\omega)\nonumber\\
&+&\Bigl[-3\partial^\mu_\omega\left(\partial^\nu_y-\partial^\nu_\omega\right)+\frac{3}{4}\eta^{\mu\nu}\partial_{x\Lambda}\left(\partial^\Lambda_y-\partial^\Lambda_\omega\right)\Bigl]\delta(\omega-x)\delta(y-x)
\eea
and
\bea
\Delta^{\mu\nu}\,_{\lambda\Lambda\delta\rho}\,G_4^{bcde, \lambda\Lambda\delta\rho}(\omega, x, y, z)&=&0.
\eea

Thus eq. (\ref{VEV2}) becomes
\bea
\bigl\langle T^{\mu\nu}(\omega)\bigl\rangle&=&-i\lim_{x,y,z\rightarrow \omega}\Biggl\{16\left(\partial^\mu_\omega\partial^\nu_x-\frac{1}{4}\eta^{\mu\nu}\partial^\alpha_\omega\partial_{x\alpha}\right)G_0(\omega-x)\nonumber\\
&+&2g^2f^{abc}f^{abc}\Delta^{\mu\nu}\,_{\lambda\delta\Lambda}\,G_3^{\lambda\delta\Lambda}(\omega, x, y)\nonumber\\
&-&i\left(n^\mu_0 n^\nu_0-n_{\alpha 0}n^\alpha_0\delta^{\mu\nu}\right)\delta^4(\omega-x')\Biggl\}.\label{eq35}
\eea

In order to include the temperature effect in the energy-momentum tensor, the TFD formalism is used and is presented briefly in the next section.

\section{Thermo field dynamics}

Thermo Field Dynamics (TFD) formalism is a quantum field theory at finite temperature \cite{Umezawa1, Umezawa2, Umezawa22, Khanna1, Khanna2}. This formalism implies that the statistical average of any operator ${\cal A}$ is interpreted as an expectation value in a thermal vacuum, i.e., $\langle {\cal A} \rangle=\langle 0(\beta)| {\cal A}|0(\beta) \rangle$. The thermal vacuum $|0(\beta) \rangle$ describes a system in thermal equilibrium, where $\beta=\frac{1}{k_BT}$, with $T$ being the temperature and $k_B$ being the Boltzmann constant. The structure of this formalism is composed of two elements: (i) doubling of the original Hilbert space and (ii) the Bogoliubov transformation.

This doubling is defined by the tilde ($^\thicksim$) conjugation rules, associating each operator in original Hilbert space, ${\cal S}$, to two operators in expanded space ${\cal S}_T$, where ${\cal S}_T={\cal S}\otimes \tilde{\cal S}$. The Bogoliubov transformation introduces the thermal effect which corresponds to a rotation in the tilde and non-tilde variables. To understand this doubling of Hilbert space let us consider 
\bea
\left( \begin{array}{cc} d(\alpha)  \\ \tilde d^\dagger(\alpha) \end{array} \right)={\cal B}(\alpha)\left( \begin{array}{cc} d(k)  \\ \tilde d^\dagger(k) \end{array} \right),
\eea
where ${\cal B}(\alpha)$ is the Bogoliubov transformation given as
\bea
{\cal B}(\alpha)=\left( \begin{array}{cc} u(\alpha) & -v(\alpha) \\ 
-v(\alpha) & u(\alpha) \end{array} \right).
\eea
Here $u(\alpha)$ and $v(\alpha)$, are related to the Bose distribution, and are given as
\bea
v^2(\alpha)=(e^{\alpha\omega}-1)^{-1}, \quad\quad u^2(\alpha)=1+v^2(\alpha).\label{phdef}
\eea

To introduce thermal effects in eq. (\ref{eq35}) the TFD formalism is applied to 2-point and 3-point Green Function, i.e., $G_0(\omega-x)$ that is the scalar field propagator, and $G_3^{\lambda\delta\Lambda}(\omega, x, y)$, that corresponds to the gluons self-interaction. 

Now consider the Minkowski space-time such that the free scalar field propagator in a doublet notation is 
\bea
G_0^{(ab)}(\omega-x;\alpha)=i\langle 0,\tilde{0}| \tau[\phi^a(\omega;\alpha)\phi^b(x;\alpha)]| 0,\tilde{0}\rangle,
\eea
where $\phi(\omega;\alpha)={\cal B}(\alpha)\phi(\omega){\cal B}^{-1}(\alpha)$ and $a, b=1,2$. Then
\bea
G_0^{(ab)}(\omega-x;\alpha)=i\int \frac{d^4k}{(2\pi)^4}e^{-ik(\omega-x)}G_0^{(ab)}(k;\alpha),
\eea
where 
\bea
G_0^{(11)}(k;\alpha)\equiv G_0(k;\alpha)=G_0(k)+v^2(k;\alpha)[G_0(k)-G^*_0(k)],
\eea
with
\bea
G_0(k)=\frac{1}{k^2-m^2+i\epsilon}
\eea
and
\bea
[G_0(k)-G^*_0(k)]=2\pi i\delta(k^2-m^2).
\eea
It is important to note that the physical information is given by the component $a=b=1$, the non-tilde component. The parameter $v^2(k;\alpha)$ in the generalized Bogoliubov transformation \cite{GBT} is given as
\bea
v^2(k;\alpha)=\sum_{s=1}^d\sum_{\lbrace\sigma_s\rbrace}2^{s-1}\sum_{l_{\sigma_1},...,l_{\sigma_s}=1}^\infty(-\eta)^{s+\sum_{r=1}^sl_{\sigma_r}}\,\exp\left[{-\sum_{j=1}^s\alpha_{\sigma_j} l_{\sigma_j} k^{\sigma_j}}\right],\label{BT}
\eea
with $d$ being the number of compactified dimensions, $\eta=1(-1)$ for fermions (bosons), $\lbrace\sigma_s\rbrace$ denotes the set of all combinations with $s$ elements and $k$ is the 4-momentum.

Similarly the 3-point function is considered, thus leading to
\bea
G_3^{(ab)\lambda\delta\Lambda}(\omega, x, y;\alpha)&=&\int\frac{d^4k_1}{(2\pi)^4}\frac{d^4k_2}{(2\pi)^4}\frac{d^4k_3}{(2\pi)^4}\,e^{-ik_1(x-y)}e^{-ik_2(x-\omega)}e^{-ik_3(\omega -y)}\nonumber\\
&\times&G_3^{(ab)\lambda\delta\Lambda}(k_1, k_2, k_3;\alpha),
\eea
where
\bea
G_3^{(ab)\lambda\delta\Lambda}(k_1, k_2, k_3;\alpha)={\cal B}^{-1(ac)}(\alpha)\,G_3^{(cd)\lambda\delta\Lambda}(k_1, k_2, k_3)\,{\cal B}^{(db)}(\alpha).
\eea
The physical component is 
\bea
G_3^{\lambda\delta\Lambda}(k_1, k_2, k_3;\alpha)&=&G_3^{\lambda\delta\Lambda}(k_1, k_2, k_3)\nonumber\\
&+&v^2(k_1, k_2, k_3; \alpha)\left[G_3^{\lambda\delta\Lambda}(k_1, k_2, k_3)-G_3^{* \lambda\delta\Lambda}(k_1, k_2, k_3)\right],
\eea
where $G_3^{\lambda\delta\Lambda}(k_1, k_2, k_3;\alpha)\equiv G_3^{(11)\lambda\delta\Lambda}(k_1, k_2, k_3;\alpha)$.

For the doubled notation, the vacuum expectation value of the energy-momentum tensor of the gluons field is, then, given by
\bea
\bigl\langle T^{\mu\nu(ab)}(\omega; \alpha)\bigl\rangle&=&-i\lim_{x,y,z\rightarrow \omega}\Biggl\{16\left(\partial^\mu_\omega\partial^\nu_x-\frac{1}{4}\eta^{\mu\nu}\partial^\alpha_\omega\partial_{x\alpha}\right)G_0^{(ab)}(\omega-x; \alpha)\nonumber\\
&+&2g^2f^{abc}f^{abc}\Delta^{\mu\nu}\,_{\lambda\delta\Lambda}\,G_3^{\lambda\delta\Lambda(ab)}(\omega, x, y; \alpha)\nonumber\\
&-&i\left(n^\mu_0 n^\nu_0-n_{\alpha 0}n^\alpha_0\delta^{\mu\nu}\right)\delta^4(\omega-x')\delta^{(ab)}\Biggl\}.
\eea

Using the standard Casimir prescription, the physical (renormalized) energy-momentum tensor is defined as
\bea
{\cal T}^{\mu\nu (ab)}(\omega;\alpha)=\langle T^{\mu\nu(ab)}(\omega;\alpha)\rangle-\langle T^{\mu\nu(ab)}(\omega)\rangle.
\eea
Explicitly this has the form
\bea
{\cal T}^{\mu\nu (ab)}(\omega;\alpha)&=&-i\lim_{x,y,z\rightarrow \omega}\Biggl\{16\left(\partial^\mu_\omega\partial^\nu_x-\frac{1}{4}\eta^{\mu\nu}\partial^\alpha_\omega\partial_{x\alpha}\right)\overline{G}_0^{(ab)}(\omega-x; \alpha)\nonumber\\
&+&2g^2f^{abc}f^{abc}\Delta^{\mu\nu}\,_{\lambda\delta\Lambda}\,\overline{G}_3^{\lambda\delta\Lambda(ab)}(\omega, x, y; \alpha)\Biggl\}\label{eq50}
\eea
where 
\bea
\overline{G}_0^{(ab)}(\omega -x;\alpha)=G_0^{(ab)}(\omega -x;\alpha)-G_0^{(ab)}(\omega -x)
\eea
and
\bea
\overline{G}_3^{\lambda\delta\Lambda(ab)}(\omega, x, y; \alpha)=G_3^{\lambda\delta\Lambda(ab)}(\omega, x, y; \alpha)-G_3^{\lambda\delta\Lambda(ab)}(\omega, x, y).
\eea

The relevant components of the Fourier representation are given by
\bea
\overline{G}_0(\omega -x;\alpha)&\equiv&\overline{G}_0^{(11)}(\omega -x;\alpha)\nonumber\\
&=&\int\frac{d^4k}{(2\pi)^4}e^{-ik(\omega -x)}v^2(k;\alpha)\left[G_0(k)-G_0^*(k)\right]
\eea
and 
\bea
\overline{G}_3^{\lambda\delta\Lambda}(\omega, x, y; \alpha)&\equiv&\overline{G}_3^{\lambda\delta\Lambda(11)}(\omega, x, y; \alpha)\nonumber\\
&=&\int\frac{d^4k_1}{(2\pi)^4}\frac{d^4k_2}{(2\pi)^4}\frac{d^4k_3}{(2\pi)^4}\,e^{-ik_1(x-y)}e^{-ik_2(x-\omega)}e^{-ik_3(\omega -y)}\nonumber\\
&\times & v^2(k_1, k_2, k_3; \alpha)\left[G_3^{\lambda\delta\Lambda}(k_1, k_2, k_3)-G_3^{* \lambda\delta\Lambda}(k_1, k_2, k_3)\right].
\eea

An important note, the eq. (\ref{eq50}) is the complete expression for the energy-momentum tensor of gluon field. The first term is the same as for the electromagnetic case multiplied by the gluons number ($n_g=8$). The second term is due to the self interactions of three gluons. The first term has been calculated \cite{review}, when the self-interaction of gluons is neglected.

In the next section, some applications are developed for different choices of $\alpha$-parameter.

\section{Stefan-Boltzmann law for gauge theory}

As a first application, consider the thermal effect that appears for $\alpha=(\beta,0,0,0)$, such that
\bea
v^2(k; \beta)=\sum_{j_0=1}^\infty e^{-\beta k j_0}
\eea
and
\bea
v^2(k_1, k_2, k_3; \beta)=\sum_{j_0=1}^\infty e^{-\beta (k_1+k_2+k_3) j_0}.
\eea
Thus the Green function becomes
\bea
\overline{G}_0^{(11)}(\omega-x;\alpha)&=&\int \frac{d^4k}{(2\pi)^4}e^{-ik(\omega-x)}\sum_{j_0=1}^\infty e^{-\beta k^0 j_0}\left[G_0(k)-G_0^*(k)\right],\nonumber\\
&=&2\sum_{j_0=1}^\infty G_0\left(\omega-x-i\beta j_0 n_0\right),\label{1GF}
\eea
where $n_0^\mu=(1,0,0,0)$. Similarly the 3-point Green function is given as
\bea
\overline{G}_3^{\lambda\delta\Lambda(11)}(\omega, x, y; \alpha)=2\sum_{j_0=1}^\infty G_3^{\lambda\delta\Lambda}\left(x-y-i\beta j_0 n_0; \omega-y-i\beta j_0 n_0; x-\omega-i\beta j_0 n_0\right).
\eea

Using $\delta'(x)=-\frac{\delta(x)}{x}$, where the prime denotes differentiation with respect to the argument, the energy-momentum tensor, eq. (\ref{eq50}), with $\mu=\nu=0$ becomes
\bea
{\cal T}^{00 (11)}(\beta)=8\left[\frac{\pi^2}{15}T^4-\left(\kappa_0 \sum_{j_0=1}^\infty\frac{\delta^2(-i\beta j_0)}{j_0^2}\right)T^2\right]
\eea
with $\kappa_0\equiv \frac{3}{2}g^2f^{abc}f^{abc}$ begin a constant and $\beta=\frac{1}{T}$ ($k_b=1$). It is important to note that the product of delta Dirac functions with identical arguments is not well defined. To avoid this problem the regularized form of delta function is defined as \cite{Van}
\bea
2\pi i\,\delta^n(x)=\left(-\frac{1}{x+i\epsilon}\right)^{n+1}-\left(-\frac{1}{x-i\epsilon}\right)^{n+1}.\label{eq60}
\eea
In order to estimate an approximated value of the self-interaction contribution, let us keep the parameter $\epsilon$ small but finite. Then considering the expansion of this term to first order is given as
\bea
\delta^2(-i\beta j_0)\approx\frac{6\epsilon}{2\pi \beta^4 j_0^4}.
\eea
 Now, using the Riemann Zeta function
\bea
\zeta(6)=\sum_{j_0=1}^\infty\frac{1}{j_0^6}=\frac{\pi^6}{945},\label{zetaf}
\eea
the 00-component of the energy-momentum tensor becomes
\bea
{\cal T}^{00 (11)}(\beta)=8\left[\frac{\pi^2}{15}T^4-\frac{\kappa_0\epsilon\pi^5}{315}T^6\right].
\eea
This is the Stefan-Boltzmann law for the gluons field. The first term is due to the free Lagrangian part and the second term corresponds to self-interaction of gluons. It is interesting to note that the gluons self-interaction has a different contribution (i.e., $\sim T^6$) to the Stefan-Boltzmann law.

\section{Casimir effect for gauge field}

In this section the Casimir effect is calculated for the non-Abelian theory at zero and finite temperature. With different choices for the $\alpha$ parameter the energy-momentum tensor is determined.

\subsection{Zero temperature}

With the choice $\alpha=(0,0,0,iL)$, the Bogoliubov transformations become
\bea
v^2(k;L)=\sum_{l_3=1}^\infty e^{-iLk l_3},
\eea
and 
\bea
v^2(k_1, k_2, k_3; L)=\sum_{l_3=1}^\infty e^{-iL (k_1+k_2+k_3) l_3}.
\eea
Then the Green function of 2-points and 3-points are given by
\bea
\overline{G}_0^{(11)}(\omega- x;L)&=&2\sum_{l_3=1}^\infty G_0\left(\omega -x-Ll_3z\right)\\
\overline{G}_3^{\lambda\delta\Lambda(11)}(\omega, x, y; L)&=&2\sum_{l_3=1}^\infty G_3^{\lambda\delta\Lambda}\left(x-y-Ll_3z; \omega-y-Ll_3z; x-\omega-Ll_3z\right).
\eea
The Casimir effect is calculated for plates separated by a distance $d$, that is related to $L = 2d$. Thus, for the component $\mu=\nu=0$ the energy-momentum tensor becomes
\bea
{\cal T}^{00 (11)}(d)=8\left[-\frac{\pi^2}{720 d^4}+\left(\frac{\kappa_0}{4} \sum_{l_3=1}^\infty\frac{\delta^2(-2dl_3)}{l_3^2}\right)\frac{1}{d^2}\right].
\eea
Using the regularized form of the delta function eq. (\ref{eq60}) and the expansion for  a small $\epsilon$, the Casimir energy is
\bea
E(d)=-\frac{8\pi^2}{720d^4}+\frac{\kappa_0\epsilon\pi^5}{2520 d^6}, 
\eea
where $E(d)\equiv {\cal T}^{00 (11)}(d)$. In similar way, the Casimir pressure is 
\bea
P(d)=-\frac{8\pi^2}{240d^4}-\frac{\kappa_0\epsilon\pi^5}{1680 d^6}, 
\eea
with $P(d)\equiv {\cal T}^{33 (11)}(d)$. The negative signs show that the Casimir force between the plates is attractive, similar to the case of the electromagnetic field.

\subsection{Results at Finite temperature}

The thermal effects are introduced by taking $\alpha=(\beta, 0, 0, i2d)$. Then the Bogoliubov transformations are
\bea
v^2(k;\beta,d)&=&v^2(k;\beta)+v^2(k;d)+2v^2(k;\beta)v^2(k;d),\nonumber\\
&=&\sum_{j_0=1}^\infty e^{-\beta kj_0}+\sum_{l_3=1}^\infty e^{-i2dkl_3}+2\sum_{j_0,l_3=1}^\infty e^{-\beta kj_0-i2dkl_3}
\eea
and
\bea
v^2(k_1, k_2, k_3;\beta,d)&=&\sum_{j_0=1}^\infty e^{-\beta (k_1+k_2+k_3) j_0}+\sum_{l_3=1}^\infty e^{-i2d (k_1+k_2+k_3) l_3}\nonumber\\
&+&2\sum_{j_0,l_3=1}^\infty e^{-\beta (k_1+k_2+k_3) j_0-i2d (k_1+k_2+k_3) l_3}.
\eea
Here the first term leads to the Stefan Boltzmann law and the second term to the Casimir effect at zero temperature. The Casimir effect at finite temperature is given by the third term.

Then the Casimir energy and pressure for the Yang-Mills field at finite temperature are
\bea
E(\beta, d)&=&{\cal T}^{00(11)}(\beta;d)\nonumber\\
&=&8\left[\frac{4}{\pi^2}\sum_{j_0,l_3=1}^\infty\frac{3(\beta j_0)^2-(2dl_3)^2}{[(\beta j_0)^2+(2dl_3)^2]^3}+\kappa_0\sum_{j_0,l_3=1}^\infty\frac{\delta^2(-i\beta j_0-2dl_3)}{(i\beta j_0+2dl_3)^2}\right],\label{ED}
\eea
and
\bea
P(\beta, d)&=&{\cal T}^{33(11)}(\beta;d)\nonumber\\
&=&8\left[\frac{4}{\pi^2}\sum_{j_0,l_3=1}^\infty\frac{(\beta j_0)^2-3(2dl_3)^2}{[(\beta j_0)^2+(2dl_3)^2]^3}-\frac{3\kappa_0}{2}\sum_{j_0,l_3=1}^\infty\frac{\delta^2(-i\beta j_0-2dl_3)}{(i\beta j_0+2dl_3)^2}\right].\label{P}
\eea
Using the regularized form of the delta function eq. (\ref{eq60}), these equations become
\bea
E(\beta, d)=8\left[\frac{4}{\pi^2}\sum_{j_0,l_3=1}^\infty\frac{3(\beta j_0)^2-(2dl_3)^2}{[(\beta j_0)^2+(2dl_3)^2]^3}+\frac{3\kappa_0}{\pi}\sum_{j_0,l_3=1}^\infty\frac{\epsilon}{(i\beta j_0+2dl_3)^6}\right],\label{ED1}
\eea
and
\bea
P(\beta, d)=8\left[\frac{4}{\pi^2}\sum_{j_0,l_3=1}^\infty\frac{(\beta j_0)^2-3(2dl_3)^2}{[(\beta j_0)^2+(2dl_3)^2]^3}-\frac{9\kappa_0}{2}\sum_{j_0,l_3=1}^\infty\frac{\epsilon}{(i\beta j_0+2dl_3)^6}\right].\label{P1}
\eea
These terms account for the simultaneous effect of spatial compactification and temperature. The contribution due to the gluon self-interaction is similar to the case at zero temperature.

\section{Conclusions}

Yang-Mills theory describes elementary particles using non-Abelian group. It is the fundamental ingredient of the unification of the electromagnetic force and weak forces as well as quantum chromodynamics, the theory of the strong force. The Stefan-Boltzmann law and the Casimir effect at zero and  finite temperature are calculated  for the non-Abelian field. The finite temperature effects are introduced using the TFD formalism, a real-time finite temperature quantum field theory. The results are composed of two parts: (i) the usual result of the electromagnetic case multiplied by the gluons number, $n_g=8$, and (ii) the contribution due to the gluons self-interaction. In the Stefan-Boltzmann law the self-interaction contribution depends of $T^6$. Then this effect has to be considered since it may be dominant at high temperature. For the Casimir energy and pressure the effect due to the self-interaction has a different dependence on the distance between the two plates ($\sim1/d^6$), i.e., the self-interaction changes the usual Casimir effect. Although the result is different the Casimir force between the plates is attractive as is in the case of the electromagnetic field. In a similar way, our results indicate that the combined effect of temperature and spatial compactification (Casimir effect at finite temperature) is modified by the self-interaction of gluons. This leads to the fact that the gluons self-interaction has to be considered at high energy.

\section*{Acknowledgments}

This work by A. F. S. is supported by CNPq projects 308611/2017-9 and 430194/2018-8.

\appendix

\section{Non-interacting massless QCD}

In this appendix the Stefan-Boltzmann law and the Casimir effect at finite temperature is studied in an approximation describing a baryon-free massless quark-gluon
plasma. Here high temperatures are considered, then the interactions and the quark mass can be discarded. The Lagrangian that describes the non-interacting massless QCD is
\bea
{\cal L}=\bar{\psi}(x)\left(i\partial_\mu\gamma^\mu\right)\psi(x)-\frac{1}{4}F_{\mu\nu}^a F^{\mu\nu a}-\frac{1}{2\alpha}\left(\partial^\mu A_\mu^a(x)\right)^2,
\eea
where $F_{\mu\nu}^a$ is the field tensor describing the gluons which is given by eq. (\ref{FF}), $\psi(x)$ is the quark field, carrying flavor and color quantum numbers and the last term is the gauge fixing term. The results at finite temperature are similar to the earlier study \cite{review}. These are presented for comparison.

In order to calculate the Casimir effect and the Stefan-Boltzmann law using the TFD formalism the energy-momentum tensor should be determine. The total energy-momentum tensor combines the quark and gluon fields, i.e.,
\bea
T_{gq}^{\mu\nu(ab)}(x)=T_g^{\mu\nu(ab)}(x)+T_q^{\mu\nu(ab)}(x).
\eea

The energy-momentum tensor for the quark field is
\bea
T_q^{\mu\nu}(x)=in_cn_f\bar{\psi}\gamma^\mu\partial^\nu\psi,
\eea
where $n_c$ and $n_f$ are the number of colors and flavors, respectively, in the $SU(3)$ non-Abelian gauge theory. In order to get the Casimir effect the energy-momentum tensor is written so as to avoid a product of field operators at the same space-time point. Then
\bea
T_q^{\mu\nu}(x)=in_cn_f\gamma^\mu \partial^\nu\lim_{x\rightarrow x'}\tau\left[\bar{\psi}(x')\psi(x)\right].
\eea

The vacuum average of the energy-momentum tensor is
\bea
\left\langle 0| T_q^{\mu\nu}(x)|0\right\rangle&=&\left\langle T_q^{\mu\nu}(x)\right\rangle\nonumber\\
&=&in_cn_f\gamma^\mu \partial^\nu\lim_{x\rightarrow x'}\left\langle 0|\tau\left[\bar{\psi}(x')\psi(x)\right]|0\right\rangle\nonumber\\
&=&-n_cn_f\lim_{x\rightarrow x'}\left\lbrace\gamma^\mu\partial^\nu S(x-x')\right\rbrace,
\eea
where the Feynman propagator for the quark (Dirac) field is
\bea
S(x-x')&=&-i\left\langle 0|\tau\left[\bar{\psi}(x')\psi(x)\right]|0\right\rangle\nonumber\\
&=&(i\gamma\cdot\partial)\,G_0(x-x'),
\eea
with $G_0(x-x')$ being the propagator of the massless scalar field. Then the energy-momentum tensor becomes
\bea
\left\langle T_q^{\mu\nu}(x)\right\rangle = -4in_cn_f\lim_{x\rightarrow x'}\left\lbrace\partial^\mu\partial^\nu G_0(x-x')\right\rbrace.
\eea

Using the TFD formalism the physical (renormalized) energy-momentum tensor is defined as
\bea
{\cal T}_q^{\mu\nu (ab)}(x;\alpha)=\langle T_q^{\mu\nu(ab)}(x;\alpha)\rangle-\langle T_q^{\mu\nu(ab)}(x)\rangle.
\eea
Then
\bea
{\cal T}_q^{\mu\nu (ab)}(x;\alpha)=-4in_cn_f\lim_{x\rightarrow x'}\left\lbrace\partial^\mu\partial^\nu\left[ G_0^{(ab)}(x-x';\alpha)-G_0^{(ab)}(x-x')\right]\right\rbrace.
\eea

The energy-momentum tensor associated to the gluon field is obtained in eq. (\ref{eq50}) as
\bea
{\cal T}_g^{\mu\nu (ab)}(x;\alpha)&=&n_g{\cal T}_{EM}^{\mu\nu (ab)}(x;\alpha)+{\cal T}_{SI}^{\mu\nu (ab)}(x;\alpha),
\eea
with $n_g$ being the gluon number, ${\cal T}_{EM}^{\mu\nu (ab)}(x;\alpha)$ being the energy-momentum tensor for the electromagnetic field, defined as
\bea
{\cal T}_{EM}^{\mu\nu (ab)}(x;\alpha)=-2i\lim_{\omega\rightarrow x}\left(\partial^\mu_\omega\partial^\nu_x-\frac{1}{4}\eta^{\mu\nu}\partial^\alpha_\omega\partial_{x\alpha}\right)\overline{G}_0^{(ab)}(\omega-x; \alpha)
\eea
and ${\cal T}_{SI}^{\mu\nu (ab)}(x;\alpha)$ being the contribution due to the gluon self-interaction given as
\bea
{\cal T}_{SI}^{\mu\nu (ab)}(x;\alpha)&=&-i\lim_{\omega,y,z\rightarrow x}\Biggl\{2g^2f^{abc}f^{abc}\Delta^{\mu\nu}\,_{\lambda\delta\Lambda}\,\overline{G}_3^{\lambda\delta\Lambda(ab)}(\omega, x, y; \alpha)\Biggl\}.
\eea
Therefore the energy-momentum tensor for the quark-gluon system is
\bea
{\cal T}_{gq}^{\mu\nu (ab)}(x;\alpha)={\cal T}_g^{\mu\nu (ab)}(x;\alpha)+{\cal T}_q^{\mu\nu (ab)}(x;\alpha).
\eea

Now the energy-momentum tensor is calculated for different choice of $\alpha$-parameter. For $\alpha=(\beta, 0, 0, 0)$ the Stefan-Boltzmann law for the quark-gluon system becomes
\bea
{\cal T}_{gq}^{00 (11)}(\beta)=8\left[\frac{\pi^2}{15}T^4-\frac{\kappa_0\epsilon\pi^5}{315}T^6\right]+n_cn_f\frac{7\pi^2}{60}T^4,
\eea
where the regularized form of the delta function eq. (\ref{eq60}) has been used in the gluon self-interaction part. 

In the case $\alpha=(0,0,0,i2d)$, the Casimir energy at zero temperature for this system is
\bea
{\cal T}_{gq}^{00 (11)}(d)=8\left[-\frac{\pi^2}{720d^4}+\frac{\kappa_0\epsilon\pi^5}{20160 d^6}\right]-n_cn_f\frac{7\pi^2}{2880 d^4}
\eea
and the Casimir pressure is
\bea
{\cal T}_{gq}^{33 (11)}(d)=8\left[-\frac{\pi^2}{240d^4}-\frac{\kappa_0\epsilon\pi^5}{13440 d^6}\right]-n_cn_f\frac{7\pi^2}{960 d^4}.
\eea

The thermal effects are exhibit for the choice $\alpha=(\beta, 0, 0, i2d)$. Then
\bea
{\cal T}_{gq}^{00 (11)}(\beta, d)&=&8\left[\frac{4}{\pi^2}\sum_{j_0,l_3=1}^\infty\frac{3(\beta j_0)^2-(2dl_3)^2}{[(\beta j_0)^2+(2dl_3)^2]^3}+\frac{3\kappa_0}{\pi}\sum_{j_0,l_3=1}^\infty\frac{\epsilon}{(i\beta j_0+2dl_3)^6}\right]\nonumber\\
&-&n_cn_f\frac{8}{\pi^2}\sum_{j_0,l_3=1}^\infty(-1)^{j_0+l_3}\frac{3(\beta j_0)^2-(2al_3)^2}{[(\beta j_0)^2+(2al_3)^2]^3}
\eea
is the Casimir energy at finite temperature and
\bea
{\cal T}_{gq}^{33 (11)}(\beta, d)&=&8\left[\frac{4}{\pi^2}\sum_{j_0,l_3=1}^\infty\frac{(\beta j_0)^2-3(2dl_3)^2}{[(\beta j_0)^2+(2dl_3)^2]^3}-\frac{9\kappa_0}{2}\sum_{j_0,l_3=1}^\infty\frac{\epsilon}{(i\beta j_0+2dl_3)^6}\right]\nonumber\\
&-&n_cn_f\frac{8}{\pi^2}\sum_{j_0,l_3=1}^\infty(-1)^{j_0+l_3}\frac{(\beta j_0)^2-3(2al_3)^2}{[(\beta j_0)^2+(2al_3)^2]^3} 
\eea
is the Casimir pressure at finite temperature.

Therefore the Stefan-Boltzmann law and the Casimir effect at zero and finite temperature for the non-interacting massless QCD have similar contributions due to free quarks and gluons and the self-interactions of gluons.

\end{document}